\documentclass[a4paper,draftcls,12pt,onecolumn]{IEEEtran}
\usepackage{mathrsfs}
\usepackage{amssymb}
\usepackage{amsfonts}
\usepackage{epstopdf}
\usepackage{graphicx}
\usepackage{caption}
\usepackage{stmaryrd}
\usepackage{amssymb,amsmath}
\usepackage{multirow,url}
\usepackage{color,cite}
\usepackage{hyperref}
\usepackage{lineno}

\ifodd 1
\newcommand{\com}[1]{\textbf{\color{red} (COMMENT: #1)}} 
\newcommand{\comg}[1]{\textbf{\color{green} (COMMENT: #1)}}
\newcommand{\response}[1]{\textbf{\color{magenta} (RESPONSE: #1)}} 
\else

\newcommand{\com}[1]{}
\newcommand{\comg}[1]{}
\newcommand{\response}[1]{}
\fi
\ifodd 0
\newcommand{\referred}[1]{\textcolor{red}{RefPaper: #1}} 
\else
\newcommand{\referred}[1]{}
\fi

\ifodd 0
\newcommand{\changeblue}[1]{\textcolor{blue}{Modified: #1}} 
\else
\newcommand{\changeblue}[1]{}
\fi

\ifCLASSINFOpdf
\else
\fi
\begin{document}

%

\title{An Optimal Decoding Strategy for Physical-layer Network Coding over Multipath Fading Channel}

\author{Minglong~Zhang,~\IEEEmembership{} Lu~Lu,~\IEEEmembership{Member,~IEEE}, and~Soung~Chang~Liew,~\IEEEmembership{Fellow,~IEEE}
\thanks{M. Zhang and L. Lu are with the Institute of Network Coding, The Chinese University of Hong Kong. S. C. Liew is with the Department of Information Engineering, The Chinese University of Hong Kong, Hong Kong.
e-mails: \{mlzhang, lulu, soung\}@ie.cuhk.edu.hk}%
\thanks{Corresponding to: lulu@ie.cuhk.edu.hk}
}




\maketitle

\begin{abstract}
We present an optimal decoder for physical-layer network coding (PNC) in multipath fading channels. Previous studies on PNC have largely focused on the single path case. For PNC, multipath not only introduces inter-symbol interference (ISI), but also cross-symbol interference (Cross-SI) between signals simultaneously transmitted by multiple users. To overcome these problems, the decoder at the relay of our PNC design makes use of a belief propagation (BP) algorithm to decode the multipath-distorted signals received from multiple users into a network-coded packet. We refer to our multipath decoding algorithm as MP-PNC. Our simulation results show that, benchmarked against synchronous PNC over a one-path channel, the bit error rate (BER) performance penalty of MP-PNC under a two-tap ITU channel model can be kept within 0.5 dB. Moreover, it outperforms a MUD-XOR algorithm by 3 dB (MUD-XOR decodes the individual information from both users explicitly before performing the XOR network-coding mapping). Although the framework of fading-channel PNC presented in this paper is demonstrated based on two-path and three-path channel models, our algorithm can be extended to cases with more than three paths.
\end{abstract}

\begin{keywords}
physical-layer network coding, multipath fading, symbol misalignment, belief propagation, asynchrony
\end{keywords}

%
%
\thispagestyle{empty}
\newpage
\setcounter{page}{1}

\section{Introduction}
%
%

\begin{figure}[h]
\centering
\captionsetup{justification=centering}
\includegraphics[width=0.8\textwidth]{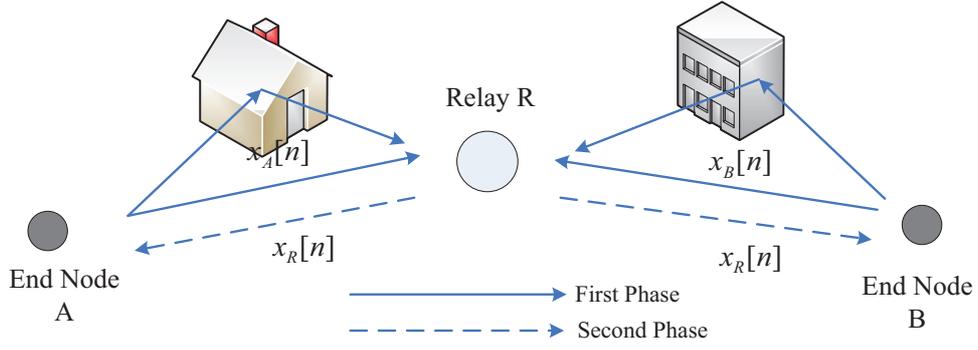}
\caption{System model for two way relay channel.} \label{fig:system}
\end{figure}

We investigate two-way relay a multipath channel where two end nodes $A$ and $B$ exchange information via a relay node $R$, as shown in Fig. \ref{fig:system}. We assume half-duplex operation and no direct channel between $A$ and $B$. A question is ``what is the minimum number of timeslots needed for the exchange of two packets between $A$ and $B$ via $R$?'' Physical-layer network coding (PNC)\cite{PNC06} requires only two time slots: one for simultaneous uplink transmissions of $A$ and $B$ to $R$, and one for broadcast downlink transmission of $R$ to $A$ and $B$. The key lies in the uplink phase, in which the relay detects the XOR of the symbols transmitted by $A$ and $B$ rather than their individual symbols.

Previous studies of PNC mostly assume the single-path fading channel. This paper considers the more general multipath fading channel. With multipath, the superposition of duplicate packets arriving at the relay node results in inter-symbol interference (ISI).

Furthermore, in \textit{asynchronous} PNC \cite {APNC_ICC} \cite{APNC_TWC}, symbols of nodes $A$ and $B$ may arrive at the relay with symbol misalignment and carrier phase offset (for both the single-path and multipath scenarios). These asynchronies between $A$ and $B$, if not properly dealt with, will lead to significant performance penalties \cite {PNCAnalysis}. Although \cite{APNC_ICC} and \cite{APNC_TWC} provided methods to reduce these performance penalties, only the single path scenario was considered. With multipath, the asynchrony problem is compounded: there are multiple symbol misalignments and carrier phase offsets between the symbols of the two transmitters. In particular, in addition to intra-user ISI, there is also Cross-SI between the two users. This paper establishes an optimal decoding framework for dealing with the ISI and Cross-SI.

\noindent \textit{\textbf{Related Work}}

Multipath and asynchrony are both pervasive in real systems. Lu and Liew \cite{APNC_ICC,APNC_TWC} proposed an optimal decoding algorithm that jointly solves the phase and symbol asynchrony problem in PNC over the AWGN channel. However, the authors of \cite{APNC_ICC} and \cite{APNC_TWC} only considered the single-path symbol-asynchronous PNC system, in which the channel realization for each end node was a flat fading (i.e., non-multipath) model.

Paper \cite{PNC_over_freq_channel} developed a decoding strategy for PNC over frequency selective channels in the time domain, but the work assumes the delays of the paths from node $A$ to relay $R$ are pairwise equal to the delays of the paths from node $B$ to relay $R$ (i.e., for each path for the former there is a corresponding path for the latter with the same delay, and vice versa). This assumption of pairwise-equal path delays is not realistic in real physical situations. We note in particular that in the multipath scenario, it is not possible to control the transmission times of the two end nodes to ensure that the signals on each and every path is aligned. Thus, for time-domain solutions, multipath PNC will necessarily be asynchronous PNC by nature.

Ref. \cite{PNC_OFDM} provided a frequency-domain OFDM solution for multipath PNC. Because the relative delay between the two collided packets (including the replicas due to multipath fading) is smaller than the cyclic prefix (CP) length, the channel is transformed to a flat fading model within each and every of the subcarrier. This effectively turns the time-domain asynchronous channel into multiple frequency-domain synchronous channels \cite{FPNC}. Although frequency-domain PNC can solve the symbol asynchrony problem, its performance is sensitive to the relative carrier frequency offset (CFO) between the two end nodes \cite{APNC_OFDM}. The CFO may cause inter-carrier interference (ICI), which may greatly degrade the system performance of a FPNC system. On the other hand, the time-domain PNC system is more sensitive to multipath fading, which may introduce inter-symbol interference (ISI) problem. This paper focuses on time-domain PNC with multipath fading.

\noindent \textit{\textbf{Contributions}}

To the best of our knowledge, no prior studies have considered the time-domain PNC over real multipath fading channels with symbol and phase asynchronies. This paper is the first to treat all the signals from multiple paths as useful information to be exploited in PNC decoding. In particular, we derive a maximum-likelihood (ML) optimal decoding algorithm based on the belief propagation (BP) algorithm that can make best use of the signals arriving from the respective multiple paths of the two users to decode and construct a network-coded packet. Extensive simulations indicate that the BER performance penalty can be kept within 0.5 dB compared with that of synchronous PNC over an additive white Gaussian noise (AWGN) channel.

\vspace{0.1in}
The remainder of this paper is organized as follows: Section \ref{Sec:sys model} describes the system model. Section \ref{Sec:jd dcod} presents our proposed optimal multipath PNC decoding algorithm. Numerical results are given in Section \ref{Sec:simulation}. Finally, Section \ref{Sec:conclusion} concludes this paper.


\section{System Model} \label{Sec:sys model}

We study a two-way relay multipath-channel network, as shown in Fig. \ref{fig:system}. Two end nodes $A$ and $B$ exchange information via a relay $R$ in the middle. We assume all nodes are half-duplex and there is no direct link between two end nodes. We adopt a two-phase transmission scheme. In this scheme, nodes $A$ and $B$ transmit uplink packets to relay $R$ simultaneously in the first phase; $R$ then constructs a network-coded downlink packet based on the collided signals and broadcasts it to both $A$ and $B$ in the second phase. After receiving the downlink packet, $A$ ($B$) can decode $B$'s packet ($A$'s packet) by subtracting its own packet (i.e., by applying the eXclusive OR (XOR) operation \cite{PNC06} with the network-coded packet).

For convenience, we express \textit {time} in units of symbol durations. That is, the duration of one symbol is 1 here.   Each symbol is carried on a rectangular pulse $g(t) = rect(t) = u(t + 1) - u(t)$.

The number of paths from  $A$ to $R$ is $p$, and the number of paths from $B$ to $R$ is $s$. Each path attenuates, delays, and introduces a phase shift to the original transmitted signals. Let $\tau _i$ be the delay of path $i$ of node $A$ and $l_j + \Delta (0 < \Delta  < 1)$ be the delay of path $j$ of node $B$. Without loss of generality, we assume $\tau _0  < \tau _1  <  \cdot  \cdot  \cdot \tau _{p - 1}$ and $l_0  < l_1  <  \cdot  \cdot  \cdot l_{s - 1}$, and we set the first path of node $A$ as the reference path and set $\tau _0  = 0$. Furthermore, we let $l_0  = 0$ so that $\Delta $ is the relative delay by which the first path of node $A$ is ahead of the first path of node $B$.

The channel impulse responses of path $i$ of node $A$ and path $j$ of node $B$ are $c_i^A (t) = \eta _i e^{-j2\pi f\tau_i} \delta(t-\tau_i)= \eta _i e^{j\varphi _i } \delta (t-\tau_i)$ and $c_j^B (t) = \mu _j e^{-j2\pi f(l_j + \Delta)}\delta(t-\Delta-l_j) = \mu _j e^{^{j\theta _j } } \delta (t-\Delta-l_j)$, respectively, where $f$ is the carrier frequency, $\eta _i$ and $\varphi _i$ ($\mu _j$ and $\theta _j$) are attenuation factors and phase shifts of path $i$ of node $A$ (path $j$ of node $B$), respectively. Then, the overall impulse response of path $i$ of node $A$, taking into consideration the pulse shape $g(t)$, is $h_i^A \left( t \right) = \int_{ - \infty }^{ + \infty } {g(\tau )} c_i^A (t - \tau ) d\tau$; similarly, the overall impulse response of path $j$ of node $B$ is $h_j^B \left( t \right) = \int_{ - \infty }^{ + \infty } {g(\tau )} c_j^B (t - \tau ) d\tau$. In other words, $h_i^A \left( t \right)$ and $h_j^B \left( t \right)$ are the effective pulse shapes.

The overall received complex baseband signal at the relay can  be expressed as
\begin{small}
\begin{align}
r(t) = &\sum\limits_{n = 1}^N \Bigg\{\sum\limits_{i = 0}^{p - 1} {x_A[n]h_i^A\left(t - (n + \tau _i )\right)} + \sum\limits_{j = 0}^{s - 1} {x_B[n]h_j^B\left(t - (n + l_j  + \Delta )\right)} \Bigg\} + w(t),
\label{eqn.1}
\end{align}
\end{small}
where $x_A [n]$ and $x_B [n]$ are the symbols of nodes $A$ and $B$, respectively, and $w(t)$ is the additive white Gaussian noise with double-sided power spectrum density $N_0 /2$.

We further assume that $0 < \tau _i  \le 1$ and $0 < l_j  \le 1 - \Delta$ for all $i,j > 0$ \footnote{We remark that if the multipath delay spread ($\tau_i$ and $l_j$) or the relative delay of two sources ($\Delta$) is larger than one symbol duration, each sample in (\ref{eqn.3}) may be embedded with more symbols. In this scenario, we can first cluster several samples with the same source symbol to create a joint symbol of higher dimensions to compute the MAP of the joint symbol. Finally we can obtain the probability for each pair of symbols $x_A[n]$ and $x_B[n]$ by doing marginalization as in (\ref{eqn.16}). Therefore, the proposed MP-PNC decoding scheme is still valid for the larger-than-one-symbol-duration multipath fading channels.}. That is, the delay spread of all paths (from $A$ as well as $B$) is within one symbol duration. This assumption of the delay spread is in accordance with some actual environments as specified in the guidelines in ITU-R M.1225 \cite{ITU_MODEL}. For example, in a three-tap channel model of an indoor office, the delay for each tap is less than 100ns. This means the assumption is suitable for a system in which the transmission rate is no more than 10Mbaud per second.

For simplicity, we first consider the case where there are only two paths between each end node and the relay (i.e., $p=2$ and $s=2$). We will show later that our method is extendable to cases with three or more paths. 
A crucial question is how relay $R$ can generate a network-coded packet from the noisy overlapped signal $r(t)$. This paper proposes a two-step decoding algorithm: 1) first oversamples  $r(t)$; 2) then use these discrete samples to build a Tanner Graph to compute the maximum \textit{a posteriori} probability (MAP) for the network-coded packet.

For 1), we consider two oversampling methods described in the following paragraphs:

\noindent \textit{\textbf{Method \uppercase\expandafter{\romannumeral1}: double sampling}}

Method \uppercase\expandafter{\romannumeral1} passes the overlapped signals $r(t)$ through two parallel matched filters and then samples their outputs at time instants $(n - 1 + \Delta )$ and $n$, $n = 1,2,...,N$, respectively. We get the following discrete-time samples:
\begin{small}
\begin{IEEEeqnarray}{Ll}
r[2n - 1] &= \frac{1}{\Delta}\int_{(n - 1)}^{(n - 1) + \Delta } {r(t)h_A^*(t-n)}dt = x_A[n]\rho _{aa}^0 + x_A[n - 1]\rho _{aa}^1 + x_B[n - 1]\rho _{ab} + w[2n - 1];\nonumber\\
r[2n] &= \frac{1}{1-\Delta}\int_{(n - 1) + \Delta }^n {r(t)h_B^* (t-n)} dt = x_B [n]\rho _{bb}^0  + x_B [n - 1]\rho _{bb}^1  + x_A [n]\rho _{ba}  + w[2n],
\label{eqn.3}
\end{IEEEeqnarray}
\end{small}
where $h_A^*(t-n) = \left(h^A _0(t-n) + h^A _1(t - \tau _1 - n)\right)^*$, $h_B^*(t) = \left(h^B _0(t - \Delta -n) + h^B _1(t - \Delta - l_1 \right. \\ \left. -n )\right)^*$ and $\rho_{aa}^i$, $\rho_{bb}^i$, $\rho_{ab}$ and $\rho_{ba} (i = 0,1)$ are integration coefficients of the corresponding matched filters. We omit the detailed expressions for the integration coefficients. Readers are referred to our technical report \cite{mp_pnc} for details.

The coefficients are all independent of $n$, since we assume the channel is unchanged during the transmission of one frame. The terms $w[2n - 1]$ and $w[2n]$ are zero-mean complex Gaussian noises with variances $\alpha_1 N_0 /2\Delta^2$ and $\alpha _2 N_0 /2(1 - \Delta)^2$, respectively, for both the real and imaginary components. Here the parameters $\alpha_1$ and $\alpha_2$ are constants given by
\begin{small}
\begin{align}
&\alpha_1 = \int_{n - 1}^{(n - 1) + \Delta } {\left| {h_A(t)} \right|^2dt};
&\alpha_2 = \int_{(n - 1) + \Delta }^{n} {\left| {h_B(t)} \right|^2dt}.
\label{eqn.5}
\end{align}
\end{small}

\noindent \textit{\textbf{Method \uppercase\expandafter{\romannumeral2}: quadruple sampling}}

Method \uppercase\expandafter{\romannumeral2} quadruples the samples. Since there are more than one paths between each end node and the relay, and each path has a different channel impulse response, we initiate a new matched filter for each path. Therefore, we adopt a fourfold sampling method at the relay. The received signal $r(t)$ serves as the input to the four different matched filters. The outputs are sampled at time instants $(n - 1 + \tau _1 )$, $(n - 1 + \Delta )$, $(n - 1 + \Delta  + l_1 )$ and $n (n = 1,2,...,N)$, accordingly. The samples are as follows:

\begin{scriptsize}
\begin{IEEEeqnarray}{Ll}
r[4n - 3] &= \frac{1}{\tau_1}\int_{(n - 1)}^{(n - 1) + \tau _1} {r(t)h^{*A} _{~0}(t - n)}dt
= x_A [n]\mu _{aa}^0 + x_A [n - 1]\mu _{aa}^1 + x_B [n - 1](\mu _{ab}^0 + \mu_{ab}^1 ) + w[4n - 3];\nonumber\\
r[4n - 2] &=\frac{1}{\Delta  - \tau _1 }\int_{(n - 1) + \tau _1 }^{(n - 1) + \Delta } {r(t)h^{*A} _{~1} (t - (n + \tau _1 ))}dt
=x_A [n](\lambda _{aa}^0  + \lambda _{aa}^1 ) + x_B [n - 1](\lambda _{ab}^0  + \lambda _{ab}^1 ) + w[4n - 2];\nonumber\\
r[4n - 1] &=\frac{1}{l_1 }\int_{(n - 1) + \Delta }^{(n - 1) + \Delta  + l_1 } {r(t)h^{*B} _{~0} (t - (n + \Delta ))}dt
=x_A [n](\mu _{ba}^0  + \mu _{ba}^1 ) + x_B [n]\mu _{bb}^0  + x_B [n - 1]\mu _{bb}^1  + w[4n - 1];\nonumber\\
r[4n] &=\frac{1} {1 - \Delta  - l_1 }\int_{(n - 1) + \Delta  + l_1 }^n {r(t)h^{*B} _{~1} (t - (n + l_1  + \Delta ))}dt
=x_A [n](\lambda _{ba}^0  + \lambda _{ba}^1 ) + x_B [n](\lambda _{bb}^0  + \lambda _{bb}^1 ) + w[4n],
\label{eqn.6}
\end{IEEEeqnarray}
\end{scriptsize}
where $\mu _{aa}^i , \mu _{ab}^i , \mu _{ba}^i , \mu _{bb}^i , \lambda _{aa}^i , \lambda _{ab}^i , \lambda _{ba}^i$ and $\lambda _{bb}^i (i = 0,1)$ are integration coefficients from the matched filters. Analogously, the terms $w[4n - 3]$, $w[4n - 2]$, $w[4n - 1]$ and $w[4n]$ are zero-mean complex Gaussian noise with variance $\beta _1 N_0 /2\tau _1^2$, $\beta _2 N_0 /2(\Delta  - \tau _1 )^2$, $\beta _3 N_0 /2l_1^2$ and $\beta _4 N_0 /2(1 - \Delta  - l_1 )^2$, respectively, for both the real and imaginary components. The parameters $\beta _1, \beta _2, \beta _3$ and $\beta _4$ are constants and can be computed like equation (\ref{eqn.5}).

\section{Joint Decoding Scheme At The Relay} \label{Sec:jd dcod}

This section presents our decoding scheme based on the BP algorithm for PNC under multipath conditions. We refer to our algorithm as MP-PNC. We assume that the relay node, by means of preambles, can perfectly estimate the channel state information (CSI), including channel impulse responses $h_i^A \left( t \right)$ and $h_j^B \left( t \right)$ , symbol timing offset $\Delta$, and propagation delays $\tau _i$ and $l_j$ in (\ref{eqn.1}). We compute the coefficients $\rho , \lambda$ and $\mu$ in (\ref{eqn.3}) and (\ref{eqn.6}). Note that the phase differences between different users and different multipath channel taps are embedded in $\rho , \lambda$ and $\mu$ already. In order to decode the joint symbol $(x_A [n],x_B [n]), n = 1,2,...,N,$ from (\ref{eqn.3}) or (\ref{eqn.6}), we need to look at $\Pr (x_A [n],x_B [n]|r_1 ,r_2 , ...,r_{2N} )$ or $\Pr (x_A [n],x_B [n]|r_1 ,r_2 , ...,r_{4N} )$, respectively. To simplify notations, let $x_A^n x_B^n $ and $r_i$ denote $(x_A [n],  x_B [n])$ and $r[i]$, respectively; and let $r$ denote $r_1 ,r_2 , ...,r_{2N}$ in (\ref{eqn.3}) or $r_1 ,r_2 , ...,r_{4N}$ in (\ref{eqn.6}). We use the BP decoding algorithm to find the exact \textit{a posteriori} probability $\Pr (x_A^n x_B^n |r)$. From this decoded probability, we can compute the maximum \textit{a posteriori} probability (MAP) of the XOR value for the downlink packet as follows:
\begin{align}
x_R [n] = \arg  \mathop {\max }\limits_x  \Pr (x_A [n] \oplus x_B [n] = x|r)
        = \arg  \mathop {\max }\limits_x  \sum\limits_{x_A^n x_B^n :x_A [n] \oplus x_B [n] = x} {\Pr (x_A^n x_B^n |r)}.
\label{equ.7}
\end{align}

We remark that the proposed MP-PNC decoding scheme is a maximum likelihood (ML) decoder, and hence optimal in terms of BER.

\subsection{Tanner graph construction}

\begin{figure}[h]
\centering
\captionsetup{justification=centering}
\includegraphics[width=0.8\textwidth]{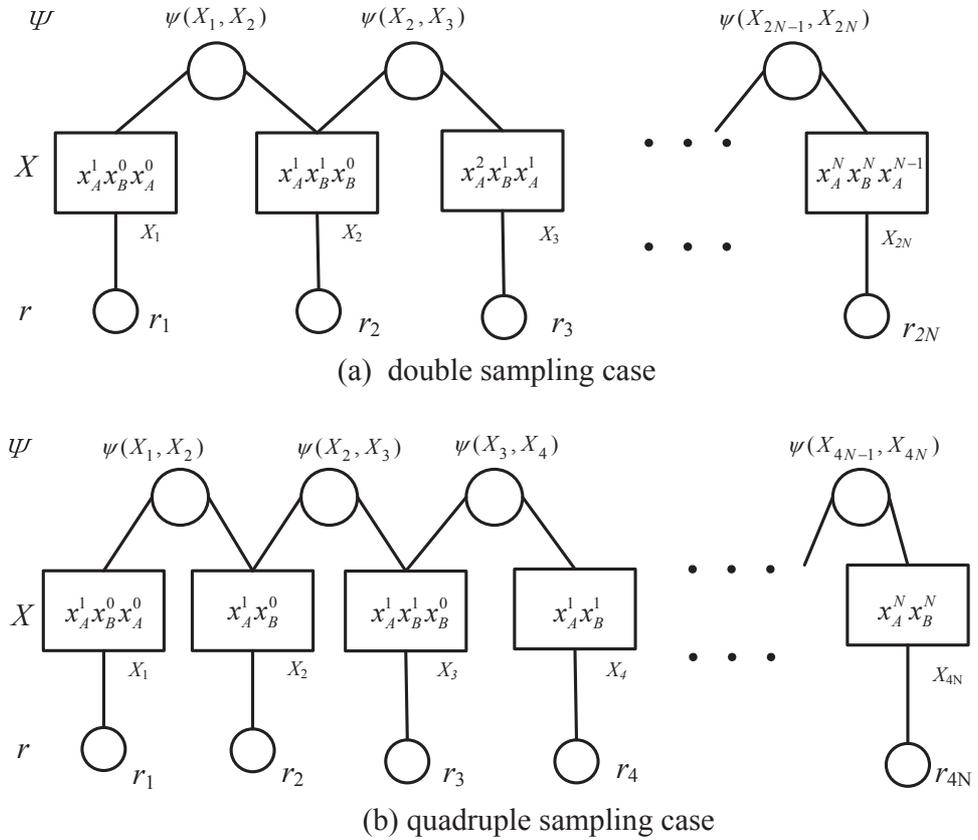}
\caption{Tanner graph constructed for MP-PNC joint decoding.} \label{fig:TannerMPPNC}
\end{figure}

Based on the relationships among the received samples in (\ref{eqn.3}) and (\ref{eqn.6}), we construct a Tanner graph as shown in Fig. \ref{fig:TannerMPPNC} (a) and Fig. \ref{fig:TannerMPPNC} (b), respectively. In Fig.\ref{fig:TannerMPPNC} (a), $X_1 , X_2 , ...,X_{2N}$ denote the 2$N$ variable nodes, and each $X_i$ is associated with a cluster of adjacent symbols from nodes $A$ and $B$ whose information is contained in sample $r_i$. Thus, $X_i$ is connected to the evidence node associated with sample $r_i$. Compatibility (or factor) nodes $\psi$ represent the connectivity among different variable nodes. Similar notations are adopted by Fig.\ref{fig:TannerMPPNC}(b) for the quadruple sampling case.

The Tanner graph in Fig. \ref{fig:TannerMPPNC} is a Markov process: e.g., $\Pr (X_i |X_{i - 1} X_{i - 2} ) = \Pr (X_i |X_{i - 1} )$. That is, given $X_{i-1}$, $X_i$ is independent of $X_{i-2}$.

For message passing from left to right, the definition of the compatibility function between variable nodes $X_{i-1}$ and $X_i$ in Fig. \ref{fig:TannerMPPNC} is
\begin{align}
\psi (X_{i - 1} ,X_i ) \propto Pr (X_{i - 1} |X_i ).
\end{align}

Our final goal is to decode the probability $\Pr (x_A^n x_B^n |r)$, from which we can obtain the ML network-coded symbol $x_R [n] = x_A [n] \oplus x_B [n]$. We first calculate $P_{2n - 1} (x_A^n x_A^{n - 1} x_B^{n - 1}  |r[2n - 1])$ and $P_{2n} (x_A^n x_B^n x_B^{n - 1} |r[2n])$ from (\ref{eqn.3}). Similarly, we compute the probability $P_{4n - 3} (x_A^n x_A^{n - 1} x_B^{n - 1} \\| {r[4n - 3]} )$, $P_{4n - 2} (x_A^n x_B^{n - 1} \left| {r[4n - 2]} \right.)$, $P_{4n - 1} (x_A^n x_B^n x_B^{n - 1} \left| {r[4n - 1]} \right.)$, and $P_{4n} (x_A^n x_B^n \left| {r[4n]} \right.)$ from (\ref{eqn.6}).


Denote the symbol set for QPSK modulation by $\chi  = \{ 1 + j, - 1 + j, - 1 - j,1 - j\}$. Assume that $a$, $b$ and $c \in \chi$, the probabilities for the evidence node $2n-1$ and $2n$ in Fig. \ref{fig:TannerMPPNC}(a) are calculated as follows:

\begin{scriptsize}
\begin{IEEEeqnarray}{Ll}
p_{2n - 1}^{a,b,c}  &= P\left( {x_A [n] = \frac{a}{\sqrt 2 },x_B [n - 1] = \frac{b}{\sqrt 2 },x_A [n - 1] = \frac{c}{\sqrt 2 }|r[2n - 1]} \right)\nonumber\\
&\propto \frac{1}{2\pi \alpha _1 \sigma ^2 /\Delta ^2}\cdot\exp \left\{ - {\frac{\left( {{\mathop{\rm Re}\nolimits} (r[2n - 1]) - {\mathop{\rm Re}\nolimits} \left(\rho_{aa}^0  \cdot a + \rho_{ab}  \cdot b + \rho_{aa}^1  \cdot c \right)/\sqrt 2 } \right)^2}{2\alpha _1 \sigma ^2 /\Delta ^2}} \right\}\cdot \nonumber\\
&~~~~~~~~~~~~~~~~~~~~~~\exp \left\{ - {\frac{\left( {{\mathop{\rm Im}\nolimits} (r[2n - 1]) - {\mathop{\rm Im}\nolimits} \left(\rho_{aa}^0  \cdot a + \rho_{ab}  \cdot b + \rho_{aa}^1  \cdot c \right)\sqrt 2 } \right)^2}{2\alpha _1 \sigma ^2 /\Delta ^2}} \right\};\nonumber
\end{IEEEeqnarray}
\begin{IEEEeqnarray}{Ll}
p_{2n}^{a,c,b}  &= P\left( {x_A [n] = \frac{a}{\sqrt 2 },x_B [n] = \frac{c}{\sqrt 2 },x_B [n - 1] = \frac{b}{\sqrt 2 }|r[2n]} \right)\nonumber\\
&\propto \frac{1}{2\pi \alpha _2 \sigma ^2 /(1-\Delta) ^2}\cdot\exp \left\{ - {\frac{\left( {{\mathop{\rm Re}\nolimits} (r[2n]) - {\mathop{\rm Re}\nolimits} \left(\rho_{ba} \cdot a + \rho_{bb}^0 \cdot c + \rho_{bb}^1  \cdot b \right)/\sqrt 2 } \right)^2}{2\alpha _2 \sigma ^2 /(1-\Delta) ^2}} \right\}\cdot \nonumber\\
&~~~~~~~~~~~~~~~~~~~~~~~~~~~~~~\exp \left\{ - {\frac{\left( {{\mathop{\rm Im}\nolimits} (r[2n]) - {\mathop{\rm Im}\nolimits} \left(\rho_{ba} \cdot a + \rho_{bb}^0 \cdot c + \rho_{bb}^1  \cdot b \right)/\sqrt 2 } \right)^2}{2\alpha _2 \sigma ^2 /(1-\Delta) ^2}} \right\} ,
\label{eqn.9}
\end{IEEEeqnarray}
\end{scriptsize}
where $\rho_{aa}^i$, $\rho_{bb}^i$, $\rho_{ab}$ and $\rho_{ba} (i = 0,1)$ are integration coefficients from the matched filters, and $\alpha_1$ and $\alpha_2$ are given in (\ref{eqn.5}). ${\mathop{\rm Re}\nolimits}(\cdot )$ and ${\mathop{\rm Im}\nolimits}(\cdot )$ denote the real and imaginary parts of the signal, respectively. Similarly, for quadruple sampling in Fig. \ref{fig:TannerMPPNC}(b), we have

\begin{scriptsize}
\begin{IEEEeqnarray}{Ll}
p_{4n - 3}^{a,b,c}  &= P\left( {x_A [n] = \frac{a}{\sqrt 2 },x_B [n - 1] = \frac{b}{\sqrt 2 },x_A [n - 1] = \frac{c}{\sqrt 2 }|r[4n - 3]} \right)\nonumber\\
&\propto \frac{1}{2\pi \beta_1 \sigma ^2 /\tau_1^2}\cdot\exp \left\{ - {\frac{\left( {{\mathop{\rm Re}\nolimits} (r[4n - 3]) - {\mathop{\rm Re}\nolimits} \left(\mu_{aa}^0  \cdot a + (\mu_{ab}^0 + \mu_{ab}^1) \cdot b + \mu_{aa}^1  \cdot c \right)/\sqrt 2 } \right)^2}{2\beta_1 \sigma ^2 /\tau_1^2}} \right\}\cdot \nonumber\\
&~~~~~~~~~~~~~~~~~~~~~\exp \left\{ - {\frac{\left( {{\mathop{\rm Im}\nolimits} (r[4n - 3]) - {\mathop{\rm Im}\nolimits} \left(\mu_{aa}^0  \cdot a + (\mu_{ab}^0 + \mu_{ab}^1) \cdot b + \mu_{aa}^1  \cdot c \right)/\sqrt 2 } \right)^2}{2\beta_1 \sigma ^2 /\tau_1^2}} \right\};\nonumber
\end{IEEEeqnarray}
\begin{IEEEeqnarray}{Ll}
p_{4n - 2}^{a,b} & = P\left( {x_A [n] = \frac{a}{\sqrt 2 },x_B [n - 1] = \frac{b}{\sqrt 2 }|r[4n - 2]} \right)\nonumber\\
&\propto \frac{1}{2\pi \beta_2 \sigma ^2 /(\Delta-\tau_1)^2}\cdot\exp \left\{ - {\frac{\left( {{\mathop{\rm Re}\nolimits} (r[4n - 2]) - {\mathop{\rm Re}\nolimits} \left((\lambda_{aa}^0 + \lambda_{aa}^1) \cdot a + (\lambda_{ab}^0 + \lambda_{ab}^1) \cdot b \right)/\sqrt 2 } \right)^2}{2\beta_2 \sigma ^2 /(\Delta-\tau_1)^2}} \right\}\cdot \nonumber\\
&~~~~~~~~~~~~~~~~~~~~~~~~~~~~~~\exp \left\{ - {\frac{\left( {{\mathop{\rm Im}\nolimits} (r[4n - 2]) - {\mathop{\rm Im}\nolimits} \left((\lambda_{aa}^0 + \lambda_{aa}^1) \cdot a + (\lambda_{ab}^0 + \lambda_{ab}^1) \cdot b \right)/\sqrt 2 } \right)^2}{2\beta_2 \sigma ^2 /(\Delta-\tau_1)^2}} \right\};\nonumber
\end{IEEEeqnarray}
\begin{IEEEeqnarray}{Ll}
p_{4n - 1}^{a,c,b} & = P\left( {x_A [n] = \frac{a}{\sqrt 2 },x_B [n] = \frac{c}{\sqrt 2 },x_B [n - 1] = \frac{b}{\sqrt 2 }|r[4n - 1]} \right)\nonumber\\
&\propto \frac{1}{2\pi \beta_3 \sigma ^2 /l_1^2}\cdot\exp \left\{ - {\frac{\left( {{\mathop{\rm Re}\nolimits} (r[4n - 1]) - {\mathop{\rm Re}\nolimits} \left((\mu_{ba}^0 + \mu_{ba}^1) \cdot a + \mu_{bb}^0 + \cdot c + \mu_{bb}^1 \cdot b \right)/\sqrt 2 } \right)^2}{2\beta_3 \sigma ^2 /l_1^2}} \right\}\cdot \nonumber\\
&~~~~~~~~~~~~~~~~~~~~\exp \left\{ - {\frac{\left( {{\mathop{\rm Im}\nolimits} (r[4n - 1]) - {\mathop{\rm Im}\nolimits} \left((\mu_{ba}^0 + \mu_{ba}^1) \cdot a + \mu_{bb}^0 + \cdot c + \mu_{bb}^1 \cdot b \right)/\sqrt 2 } \right)^2}{2\beta_3 \sigma ^2 /l_1^2}} \right\};\nonumber
\end{IEEEeqnarray}
\begin{IEEEeqnarray}{Ll}
p_{4n}^{a,c} & = P\left( {x_A [n] = \frac{a}{\sqrt 2 },x_B [n] = \frac{c}{\sqrt 2 }|r[4n]} \right)\nonumber\\
&\propto \frac{1}{2\pi \beta_4 \sigma ^2 /(1-\Delta-\tau_1)^2}\cdot\exp \left\{ - {\frac{\left( {{\mathop{\rm Re}\nolimits} (r[4n]) - {\mathop{\rm Re}\nolimits} \left((\lambda_{ba}^0 + \lambda_{ba}^1) \cdot a + (\lambda_{bb}^0 + \lambda_{bb}^1) \cdot c \right)/\sqrt 2 } \right)^2}{2\beta_4 \sigma ^2 /(1-\Delta-\tau_1)^2}} \right\}\cdot \nonumber\\
&~~~~~~~~~~~~~~~~~~~~~~~~~~~~~~~~~~~~\exp \left\{ - {\frac{\left( {{\mathop{\rm Im}\nolimits} (r[4n]) - {\mathop{\rm Im}\nolimits} \left((\lambda_{ba}^0 + \lambda_{ba}^1) \cdot a + (\lambda_{bb}^0 + \lambda_{bb}^1) \cdot c + \right)/\sqrt 2 } \right)^2}{2\beta_4 \sigma ^2 /(1-\Delta-\tau_1)^2}} \right\}.
\label{eqn.10}
\end{IEEEeqnarray}
\end{scriptsize}

\subsection{Message update rules}

\begin{figure}[h]
\centering
\captionsetup{justification=centering}
\includegraphics[width=0.8\textwidth]{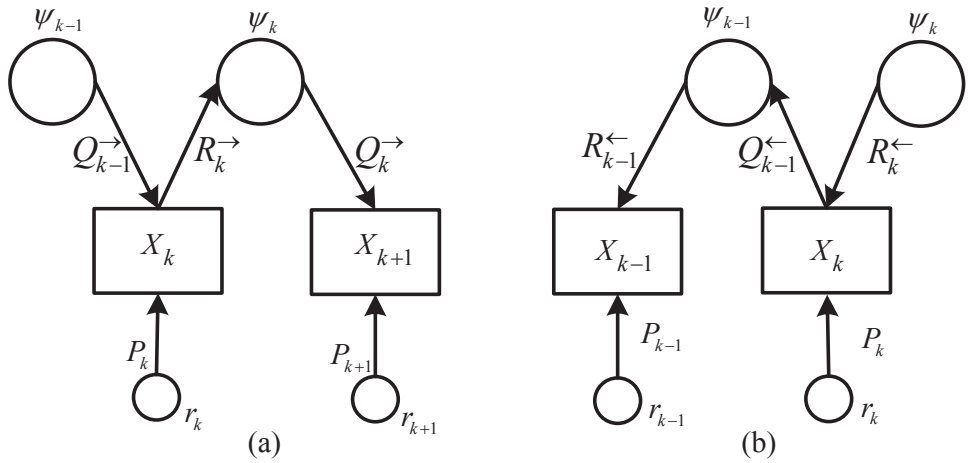}
\caption{(a)From left to right; ~~~~~~~~(b)From right to left. \protect\\ Message updating rules in Tanner graph.} \label{fig:update}
\end{figure}

We make use of the message from each evidence node as in (\ref{eqn.9}) and (\ref{eqn.10}) to derive the message update rules for the Tanner graph in Fig. \ref{fig:update}. The Tanner graph has a tree structure, implying only one iteration is enough (one message update on each edge) for convergence of the algorithm. We update the right-bound messages from left to right, and then the left-bound messages  from right to left, as illustrated in Fig. \ref{fig:update}.
In Fig. \ref{fig:update}, for the double sampling case, $Q_k$ and $R_k$ denote the right-bound and left-bound messages on the edge of the \textit{k}-th compatibility node, respectively. $P_k = (p_k^{1 + j,1 + j,1 + j} ,p_k^{1 + j,1 + j, - 1 + j} ,...,p_k^{1 - j, - 1 - j,1 - j} )$ is a $64 \times 1$ probability vector, where each component is the joint conditional probability $p_k^{a,b,c}$ in (\ref{eqn.9}). Similarly, $Q_{k - 1} = (q_{k - 1}^{1 + j,1 + j,1 + j}, q_{k - 1}^{1 + j,1 + j, - 1 + j},..., q_{k - 1}^{1 - j, - 1 - j,1 - j} )$ and $R_k = (r_k^{1 + j,1 + j,1 + j} ,r_k^{1 + j,1 + j, - 1 + j} ,...,r_k^{1 - j, - 1 - j,1 - j} )$ are also $64 \times 1$ probability vectors where $q_{k - 1}^{a,b,c}$ and $r_{k}^{a,b,c}$ are probabilities $P(x_A [\left\lceil {k/2} \right\rceil ] = a,x_B [\lceil {k/2} \rceil ] = c,x_B [\left\lfloor {k - 1/2} \right\rfloor ] = b|r[1],...,r[k - 1])$ and $P(x_A [\left\lceil {k/2} \right\rceil ] = a,x_B [\left\lfloor {k-1/2} \right\rfloor ] = b,x_A [\left\lfloor {k-1/2} \right\rfloor ] = c|r[1],...,r[k])$, respectively. Note that in Fig. \ref{fig:update}, for $Q_k$ and $R_k$, we have an arrowhead $\to$ for the right-bound messages $Q_k^\to$ and $R_k^\to$ and an arrowhead $\leftarrow$ for the left-bound messages $Q_k^\leftarrow$ and $R_k^\leftarrow$. The right-bound and left-bound messages are distinct and not the same.

According to the principle of the BP algorithm (also known as the sum-product algorithm), the output of a node should be consistent with all its inputs when summing over the products of all possible input combinations \cite{BP}. For our Tanner graph, the details are as follows:

\noindent \textbf{1) Update of right-bound messages}

With reference to Fig. \ref{fig:update} (a), suppose that we want to update $Q_k^\to$ from $P_k$ and $Q_{k-1}^\to$.  Based on the sum-product principle, for each element $r_k^{a,b,c}$ in $Q_k^\to$, we compute
\begin{small}
\begin{align}
&r_k^{a,b,c}  = p_k^{a,b,c} \cdot q_{k - 1}^{a,b,c}.
\label{eqn.11}
\end{align}
\end{small}
from $p_k^{a,b,c}$ and $q_{k-1}^{a,b,c}$ in $P_k$ and $Q_{k-1}^\to$, respectively. For the input message going into the leftmost compatibility node, (\ref{eqn.11}) should be $r_k^{a,b,c}  = p_k^{a,b,c}$.

To update the message $Q_k^\to$ from $R_k^\to$, note that $Q_k^\to$ is from compatibility node $\psi _k$ and $R_k^\to$ is from variable node $X_k$. Suppose that for $X_k$ and $X_{k+1}$, the common symbols overlapped in the two adjacent samples are $a$ and $c$. Then we have
\begin{small}
\begin{align}
&q_k^{a,1 + j,c}  = q_k^{a, - 1 + j,c} = q_k^{a, - 1 - j,c}  = q_k^{a,1 - j,c}  = \sum\limits_b {r_k^{a,b,c} }.
\label{equ.12}
\end{align}
\end{small}
Similarly, if the common symbols are b and c, (or a and b), the update equation are
\begin{small}
\begin{align}
&q_k^{1 + j,b,c}  = q_k^{ - 1 + j,b,c} = q_k^{ - 1 - j,b,c}  = q_k^{1 - j,b,c}  = \sum\limits_a {r_k^{a,b,c} } ;
\label{equ.13}
\end{align}
\end{small}
or
\begin{small}
\begin{align}
&q_k^{a,b,1 + j}  = q_k^{a,b, - 1 + j}  = q_k^{a,b, - 1 - j}  = q_k^{a,b,1 - j}  = \sum\limits_c {r_k^{a,b,c} }.
\label{equ.14}
\end{align}
\end{small}
By applying the update rules described in (\ref{eqn.11})-(\ref{equ.14}), we can update the next message $R_{k+1}^\to$ and $Q_{k+1}^\to$, and so on and so forth until we reach the right-most node.

\noindent \textbf{2) Update of left-bound messages}

With reference to Fig. \ref{fig:update}(b), we use a similar method as in 1) to update the left-bounded messages. Moreover, for the quadruple sampling case, the message passing procedure is analogous to the double sampling case discussed.

\subsection{Decision Making}

After the message passing process, for the double sampling case, at each even evidence node we have
\begin{small}
\begin{align}
&p(x_A [n] = a,x_B [n] = b,x_B [n - 1] = c|r) = \mu (X_{2n} ) = p_{2n}^{a,b,c}  \cdot q_{2n - 1}^{a,b,c}  \cdot r_{2n}^{a,b,c}.
\label{eqn.15}
\end{align}
\end{small}
For the last node, (\ref{eqn.15}) is modified by omitting $r_{2N}^{a,b,c}$. By marginalizing the variable $x_B [n - 1]$, we compute the ML network-coded symbol:
\begin{small}
\begin{align}
&x_R [n] = x_A [n] \oplus x_B [n] = \arg \mathop {\max }\limits_{x \in \chi } \left(\sum\limits_{x = x_A [n] \oplus x_B [n]} {\sum\limits_{  x_B [n - 1]} {\mu (X_{2n} )} } \right).
\label{eqn.16}
\end{align}
\end{small}
For the quadruple sampling case, the decision making equation is analogous:
\begin{small}
\begin{align}
&x_R [n] = x_A [n] \oplus x_B [n] = \arg  \mathop {\max }\limits_x \left(\sum\limits_{x = x_A [n] \oplus x_B [n]} {\mu (X_{4n} } ) \right).
\label{eqn.17}
\end{align}
\end{small}
where $\mu (X_{4n} ) = p_{4n}^{a,b}  \cdot q_{4n - 2}^{a,b} \cdot r_{4n - 1}^{a,b}$. Note that for every fourth variable node in Fig. \ref{fig:TannerMPPNC} (b), there are only two variables: $x_A [n]$ and $x_B [n]$. Therefore, we do not need to do marginalization as in (\ref{eqn.16}).

\subsection{Extension to Multipath Channel with more than Two Paths}

The decoding algorithm presented in the above subsections is not only suitable for the two-tap channel model, but can also be easily extended to the multiple-tap (i.e., more than two paths) channel model.  Specifically, if the number of paths is three and the last tap arrival time is still within the first symbol duration, a Tanner graph can be derived in a similar manner. For instance, we can still adopt the double sampling method to construct a Tanner graph, and then update the messages. Let us take QPSK modulation for an example, the combination turns out to contain four variables $(x_A [n],x_A [n - 1],x_B [n],x_B [n - 1])(n = 1,2,...,N)$ at most in (\ref{eqn.9}) and becomes a $256 \times 1$ probability vector. Therefore, the complexity of the update rules is no more than 256 multiplications (see (\ref{eqn.11})), and the other operations are simple additions. Similarly, we can extend our method to the four or more tap channel model. However, the computation complexity will increase quickly. In practice, in an indoor application scenario, we typically need to consider only the first three paths because the energy for the fourth path and thereafter decays drastically (their overall power is less than 1\% to the total reception power \cite{ITU_MODEL}).

\section{Simulation Results} \label{Sec:simulation}

In this section, we present the numerical simulation results for MP-PNC. The synchronous PNC without multipath \cite{APNC_ICC} and an extended MUD-XOR decoding method are used as benchmarks for evaluating the average bit error rate (BER) performance. The QPSK symbol amplitude is scaled to $\sqrt 2$ of the BPSK symbol amplitude to equalize the per-bit energy.  Moreover, for  fair comparison, we equalize the per-bit SNRs in the multipath system and the AWGN single-path system.

\subsection{Channel Model}

In our simulation, we adopt the empirical multipath channel model, specified in the ITU-R M.1225 \cite{ITU_MODEL}. In particular, we choose two different three-tap indoor office channel models as the wireless channels between two end nodes and the relay, respectively. The channel impulse responses between node $A$ (node $B$) and relay R are as follows:
\begin{small}
\begin{align}
&h_A (t) = \delta (t) + 0.7079\delta (t - 0.05) + 0.3162\delta (t - 0.11)\nonumber\\
&h_B (t) = \delta (t) + 0.6808\delta (t - 0.1) + 0.4365\delta (t - 0.2),
\label{eqn.18}
\end{align}
\end{small}
where we assume the bandwidth for each channel is 1 MHz. The amplitude of the first tap is normalized 1, and the largest delay spread (i.e., the last path) is within one symbol duration. As discussed in Section \ref{Sec:jd dcod}, the signal power for the fourth path and thereafter is very weak, and therefore we omit them in the simulation.

\subsection{Extended Disjoint MUD-XOR Decoding Scheme}

The MP-PNC is a joint decoding algorithm. To benchmark MP-PNC,  we consider a disjoint MUD-XOR decoding scheme that decodes $x_A [n]$ and $x_B [n](n = 1,2,...,N)$ individually before XORing them. For MUD-XOR here we extend a previous single-path decoding algorithm \cite{CRESM} to a multipath one. The extended disjoint MUD-XOR decoding scheme is elaborated in our technical report \cite{mp_pnc} and we omit the details here. In the extended disjoint MUD-XOR, the relay $R$ constructs a downlink network-coded symbol $x_R [n] = x_A [n]  \oplus  x_B [n]$, based on the decoded individual symbols of $x_A [n]$ and $x_B[n]$ from two end nodes.

\subsection{BER Performance Evaluation}

Let $L$ denote the number of paths in the uplink channel of node $A$ and $B$, and $\phi^A=\phi_0 ^A,\phi_1 ^A ...$ and $\phi^B=\phi_0 ^B,\phi_1 ^B ...$ denote the relative phase rotations of the other paths relative to the phase rotation of the first path of node $A$ (i.e., $\phi_0 ^A = 0$). Note that the phase rotations $\phi_i ^A$ and $\phi_j ^B$ are the effective phase rotations, which may be caused by the path delays, the reflections during propagation and so on. The phase terms $\phi_i ^A$ ($\phi_j ^B$) are different from $\varphi _i$ ($\theta _j$) defined in the channel impulse responses in Section \ref{Sec:sys model}. Fig. \ref{fig:two_tap} plots the BER performance of the relay node using BPSK and QPSK modulations for both the double and quadruple sampling methods. The x-axis is the average per-bit SNR (its unit is dB) of the two paths and the y-axis is the average BER. In order to study the performance of the proposed MP-PNC algorithm, we choose synchronous PNC over a single-path AWGN channel and the extended disjoint MUD-XOR scheme with the same multipath channel (see (\ref{eqn.18})) as benchmarks. From the Fig. \ref{fig:two_tap}, we can see that compared with synchronous PNC (whose BER serves as lower inner-bound), the performance penalty of the MP-PNC decoding algorithm in a two-tap channel is only approximately 0.5 dB. Moreover, it outperforms the MUD-XOR scheme by 3 dB. In addition, quadruple sampling leads to better performance than double sampling. Note that we use a rectangular pulse shaping function, which is not band-limited. Therefore, quadruple sampling beyond the Nyquist rate will provide more information for computing \textit{a posteriori} probability. Fig. \ref{fig:three_tap} provides BER performance for the three-tap channels. We use the double sampling method in this scenario. The gap between the synchronous case and the multipath case is less than 1 dB. We can also see that the BP-based algorithm has much better performance than that of the MUD-XOR scheme.

We plot the impact of different symbol misalignments on the BER performance in Fig. \ref{fig:diff_misalignments}. We find that the optimal value of $\Delta$ is 0.5, for both the BPSK and the QPSK modulations. The relationship between the BER values and the symbol misalignments $\Delta$ is elaborated in \cite{mp_pnc}.

Fig. \ref{fig:diff_phase} shows the impact of different phase rotations on the BER performance of our proposed MP-PNC over two-tap channels. The constellation map of the received signal is changed by the phase rotations. The BER depends on the relative phase rotations of all paths, not just one or two of them. For BPSK, we find that there is little performance difference between the case with large relative phase rotations (say $\phi ^A = 0, \pi/10$ and $\phi ^B = 5\pi/6, 3\pi/4$) and the case with small relative phase rotations (say $\phi ^A = 0, \pi/10$ and $\phi ^B = \pi/8, \pi/6$) between nodes $A$ and $B$. For QPSK, benchmarked against the case with small relative phase rotations (say $\phi ^A = 0, \pi/10$ and $\phi ^B = \pi/8$, $\pi/6$) between two end nodes, there is a very small performance penalty (less than 0.3 dB) if the relative phase rotations are $\phi ^A = 0, \pi/10$ and $\phi ^B = \pi/8$, $2\pi/3$ (also valid if $\phi ^A = 0, \pi/3$ and $\phi ^B = \pi/8, \pi/6$). However, the performance penalty can be as large as 1 dB (or 2 dB) if the relative phase rotations are $\phi ^A = 0, \pi/10$ and $\phi ^B = \pi/2$, $2\pi/3$ (or $\phi ^A = 0, \pi/10$ and $\phi ^B = 2\pi/3$, $\pi/6$). Based on the above observations, we can reach a conclusion that, for the channel model we adopted, although the BER performance of MP-PNC depends on the channel gains of all paths, the relative phase rotation between two strongest paths (e.g., the first paths of the two end nodes in our channel model) has a larger impact on the BER.

Although the above conclusions are obtained by applying the channel realization specified by (\ref{eqn.18}), the proposed MP-PNC decoding algorithm is valid for any multipath fading channel model. We have verified these results with a completely different channel model as shown in (\ref{eqn.19}) and all the simulation results are shown in Fig. \ref{fig:two_tap2} -- Fig. \ref{fig:diff_phase2}. In general, the observations on the BER performance of our MP-PNC method for the ITU channel model (\ref{eqn.18}) and the new channel of (\ref{eqn.19}) models are similar except for the last observation. For the last observation, we have concluded that the relative phase difference between the strongest paths of the two users has a larger impact on the BER performance, compared with the phase difference between the other paths. However, from Fig. 12, we can see that the phase difference between the two secondary paths in the new model also has non-negligible impact on the BER performance. The reason is that the power difference between the first and the second channel taps in the new channel model in (\ref{eqn.19}) is relatively small, hence we could not ignore the effect of the secondary channel taps.
\begin{small}
\begin{align}
&h_A (t) = \delta (t) + 0.9487\delta (t - 0.15) + 0.3162\delta (t - 0.25)\nonumber\\
&h_B (t) = \delta (t) + 0.9644\delta (t - 0.35) + 0.3873\delta (t - 0.45).
\label{eqn.19}
\end{align}
\end{small}


\section{Conclusion And Future Work} \label{Sec:conclusion}

This paper has proposed an optimal maximum-likelihood (ML) decoder for physical-layer network coding (PNC) over multipath fading channel, referred to as MP-PNC, in a two-way relay network. The decoding algorithm is based on belief propagation (BP). Instead of regarding the signals from non-major paths as interferences, MP-PNC can fully exploit the signals from non-major channel paths for decoding. Therefore, it effectively improves the BER performance of the multipath system, compared with disjoint MUD-XOR decoding algorithm. Specifically, simulation results show that with BP decoding, the symbol asynchrony and multipath fading issues can be solved in an integrated manner.

This work has only studied non-channel-coded PNC systems. Going forward, the study of channel-coded PNC under the multipath fading scenario will be interesting. In addition, investigation of PNC via a large delay multipath fading channel is also worthwhile.


\section*{Acknowledgment}
This work is partially supported by the General Research Funds (Project No. 414713) and AoE grant E-02/08, established under the University Grant Committee of the Hong Kong Special Administrative Region, China. This work is also partially supported by the China NSFC grant (Project No. 61271277).

{\scriptsize
\bibliographystyle{IEEEtran}
\bibliography{mppnc}
}

\pagebreak

\begin{figure}[h]
\centering
\includegraphics[width=0.7\textwidth]{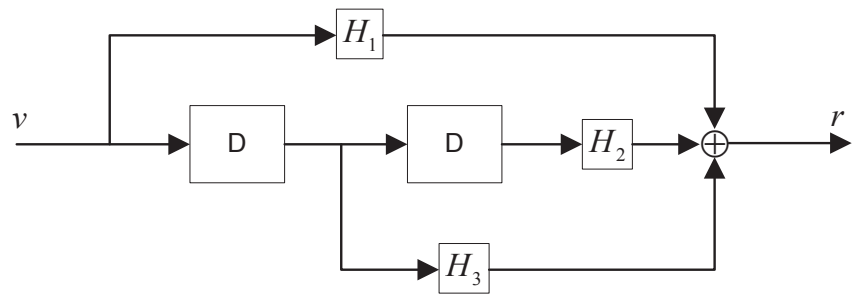}
\caption{The virtual encoder for MUD-XOR system, where D stands for one symbol delay.} \label{fig:mud-xor}
\end{figure}

\begin{figure}[h]
\centering
\includegraphics[width=1\textwidth]{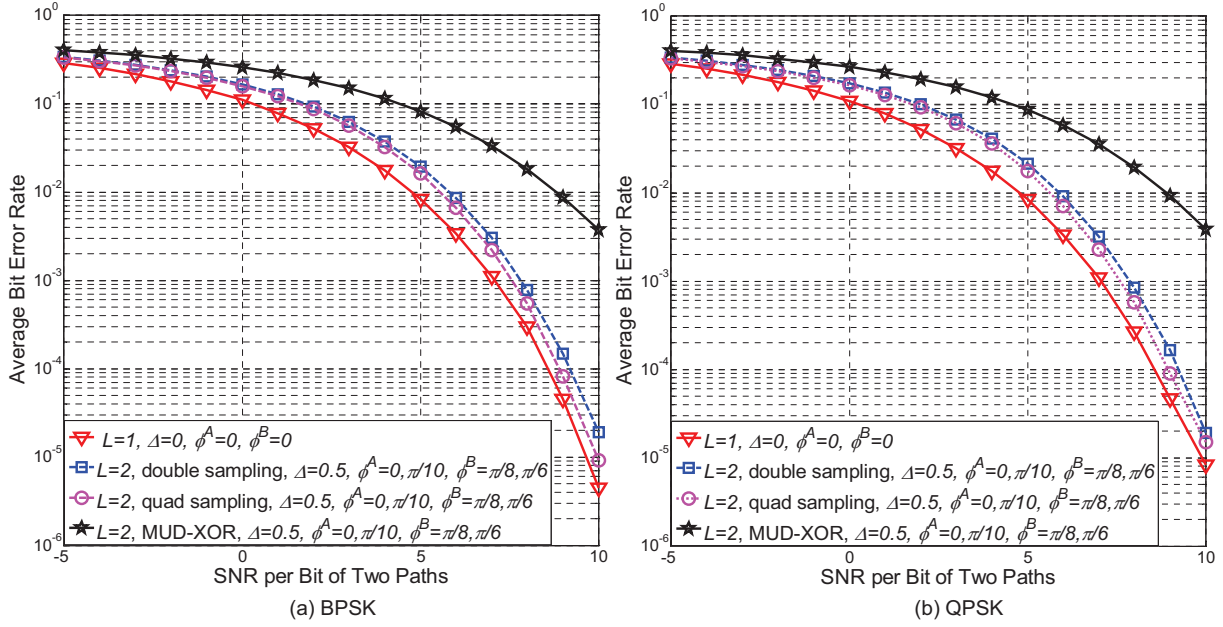}
\caption{BER curves for MP-PNC decoding in a two-tap multipath channel: (a) BPSK; (b) QPSK. $L$ denotes the number of paths, $\Delta$ denotes the symbol offset between the two first paths from the two end nodes, and $\phi$ represents the relative phase rotation of other paths with respect to the first path of end node $A$.} \label{fig:two_tap}
\end{figure}

\begin{figure}[h]
\centering
\captionsetup{justification=centering}
\includegraphics[width=1\textwidth]{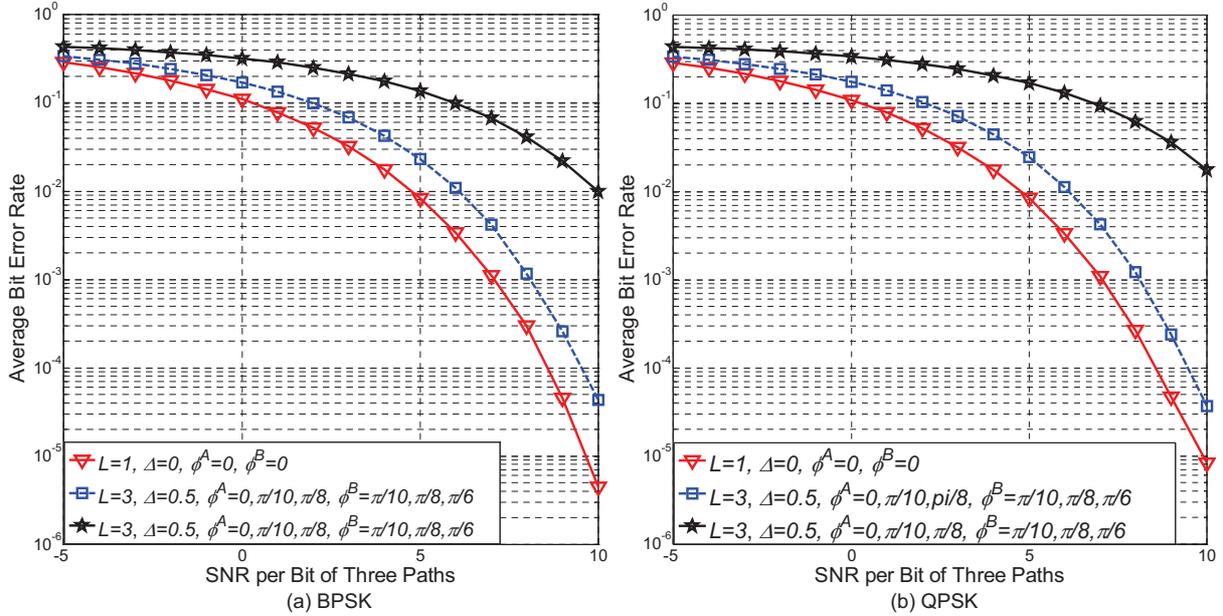}
\caption{PNC via a three-tap multipath channel.} \label{fig:three_tap}
\end{figure}

\begin{figure}[h]
\centering
\captionsetup{justification=centering}
\includegraphics[width=1\textwidth]{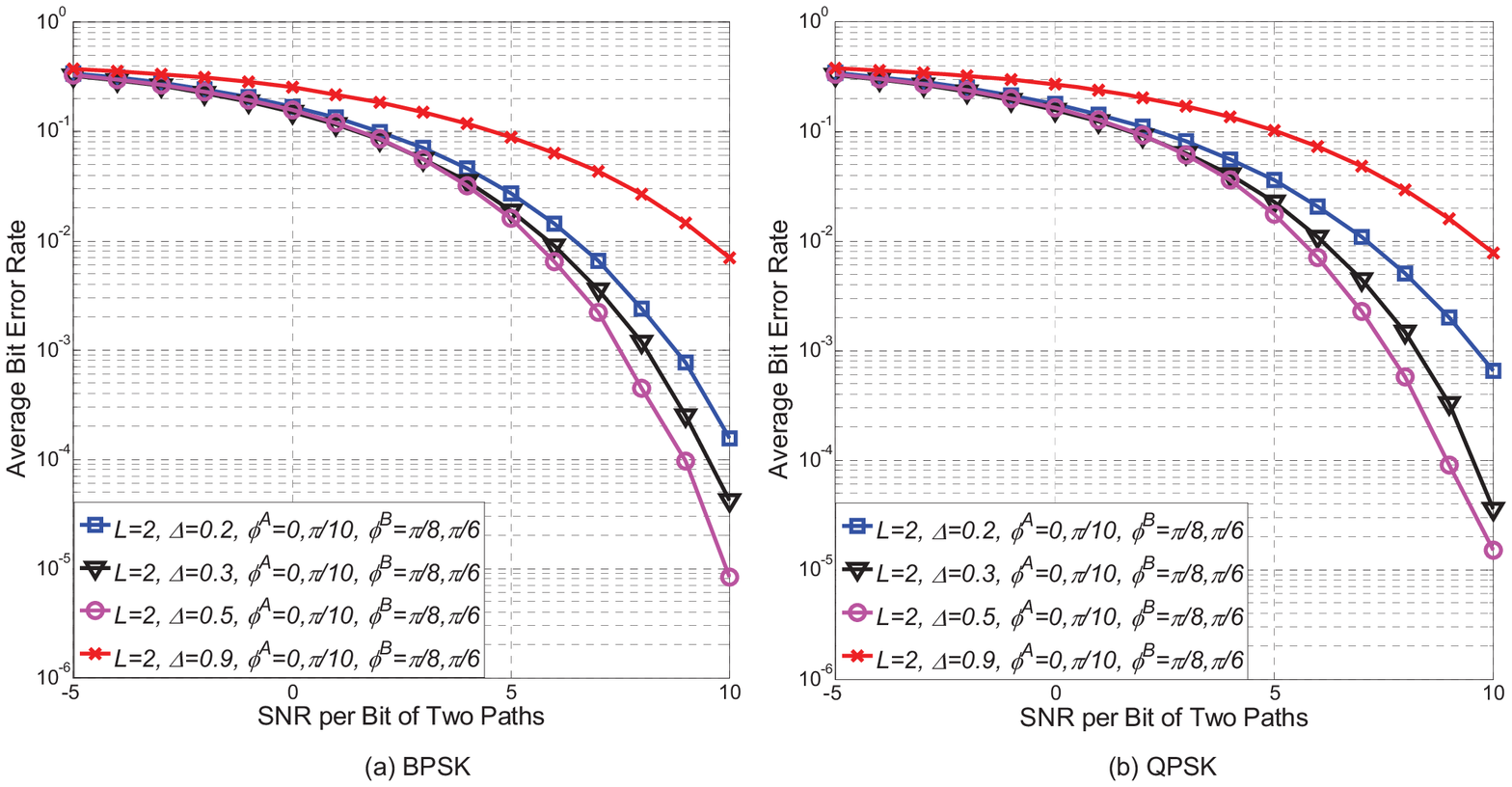}
\caption{BER performance with different symbol misalignments.} \label{fig:diff_misalignments}
\end{figure}

\begin{figure}[h]
\centering
\includegraphics[width=1\textwidth]{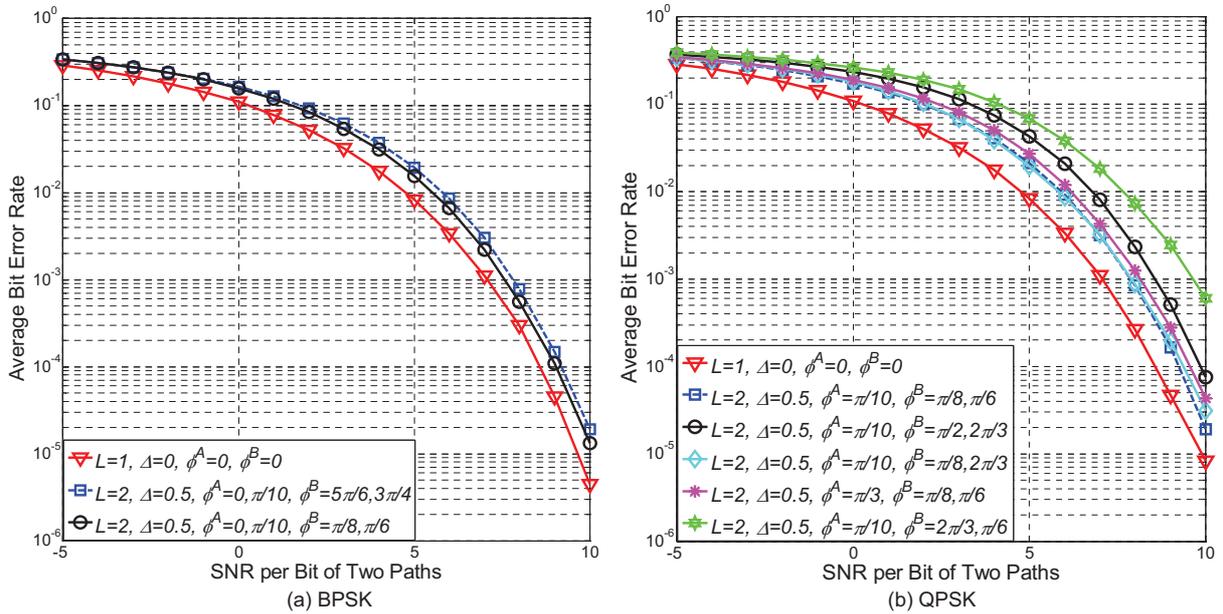}
\caption{BER performance with different phase rotations: for BPSK, the performance difference between the case with large phase rotations and the case with small phase rotations is very small; for QPSK, the relative phase rotations between two strongest paths has a larger impact on the BER performance.} \label{fig:diff_phase}
\end{figure}

\begin{figure}[h]
\centering
\includegraphics[width=1\textwidth]{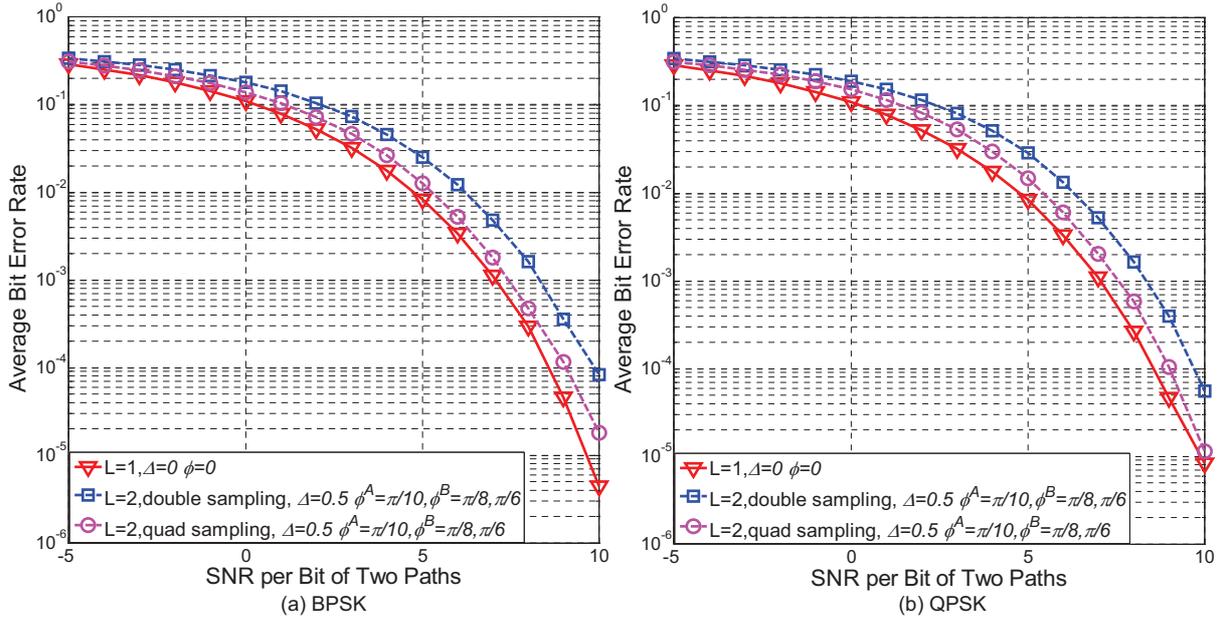}
\caption{BER curves for MP-PNC under the channel of (\ref{eqn.19}), in which we only consider the first two taps for simplicity of simulation.  The term $L$ denotes the number of paths, $\Delta$ denotes the symbol offset between the two first paths from the two end nodes, and $\phi^A$ and $\phi^B$ represent the relative phase rotations of the other paths with respect to the first path of node A.} \label{fig:two_tap2}
\end{figure}

\begin{figure}[h]
\centering
\captionsetup{justification=centering}
\includegraphics[width=1\textwidth]{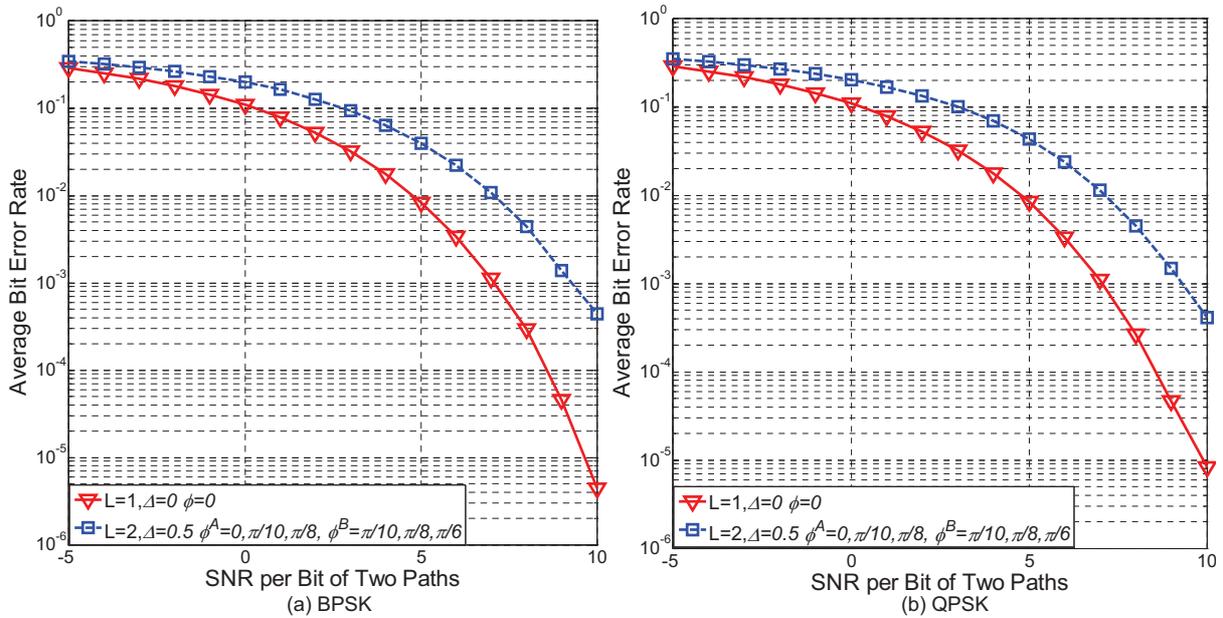}
\caption{MP-PNC via the three-tap channel as described in (\ref{eqn.19}).} \label{fig:three_tap2}
\end{figure}

\begin{figure}[h]
\centering
\captionsetup{justification=centering}
\includegraphics[width=1\textwidth]{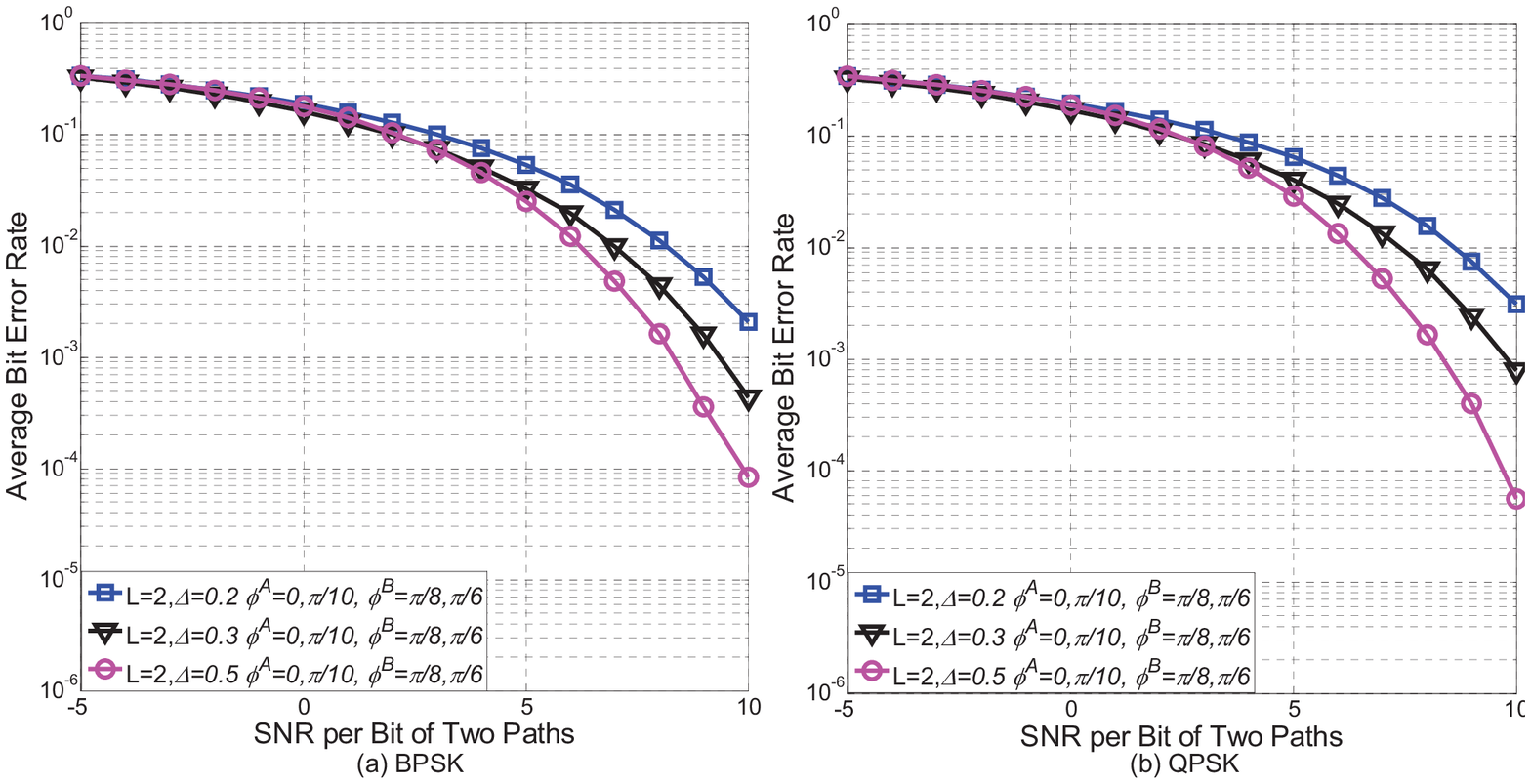}
\caption{BER performance with different symbol misalignments.} \label{fig:diff_misalignments2}
\end{figure}

\begin{figure}[h]
\centering
\includegraphics[width=1\textwidth]{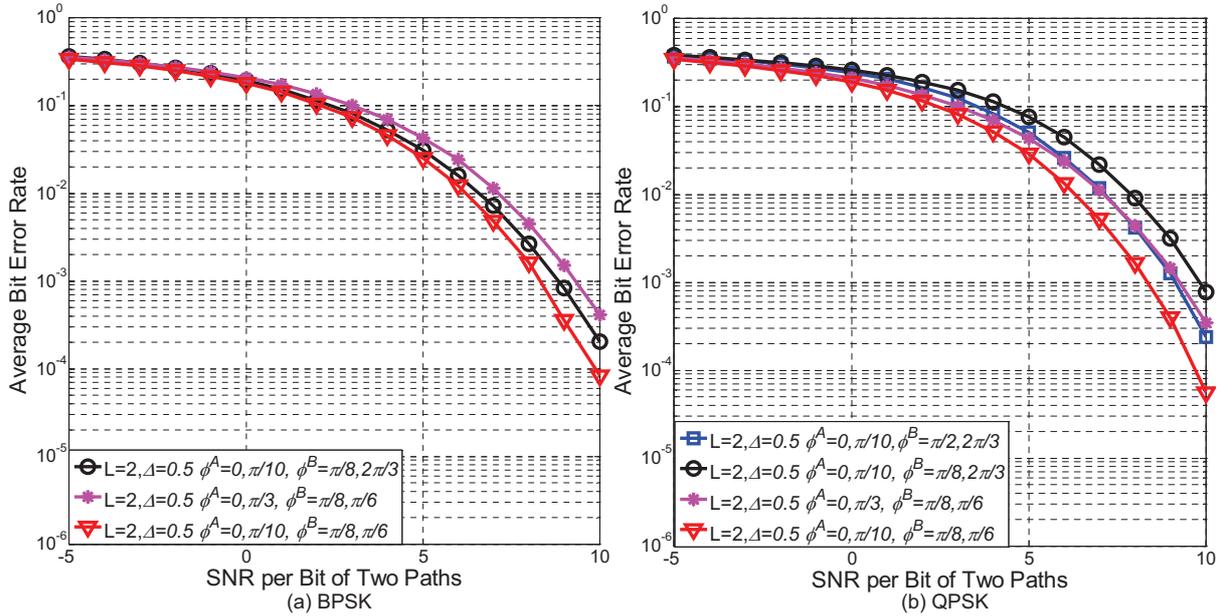}
\caption{BER performance with different phase rotations. Compared with the conclusions for Fig. \ref{fig:diff_phase} above, for QPSK, the relative phase difference between two second paths, whose powers are comparable to the main path, could have a non-negligible impact on the BER performance (e.g., comparing the black curve with the red curve).} \label{fig:diff_phase2}
\end{figure}
\end{document}